\documentclass[]{article}
\usepackage{lmodern}
\usepackage{amssymb,amsmath}
\usepackage{ifxetex,ifluatex}
\ifnum 0\ifxetex 1\fi\ifluatex 1\fi=0 
  \usepackage[T1]{fontenc}
  \usepackage[utf8]{inputenc}
  \usepackage{textcomp} 
\else 
  \usepackage{unicode-math}
  \defaultfontfeatures{Scale=MatchLowercase}
  \defaultfontfeatures[\rmfamily]{Ligatures=TeX,Scale=1}
\fi
\IfFileExists{upquote.sty}{\usepackage{upquote}}{}
\IfFileExists{microtype.sty}{
  \usepackage[]{microtype}
  \UseMicrotypeSet[protrusion]{basicmath} 
}{}
\makeatletter
\@ifundefined{KOMAClassName}{
  \IfFileExists{parskip.sty}{%
    \usepackage{parskip}
  }{
    \setlength{\parindent}{0pt}
    \setlength{\parskip}{6pt plus 2pt minus 1pt}}
}{
  \KOMAoptions{parskip=half}}
\makeatother
\usepackage{xcolor}
\IfFileExists{xurl.sty}{\usepackage{xurl}}{} 
\IfFileExists{bookmark.sty}{\usepackage{bookmark}}{\usepackage{hyperref}}
\hypersetup{
  pdftitle={The Unintended Consequences of Censoring Digital Technology -- Evidence from Italy's ChatGPT Ban},
  pdfauthor={David Kreitmeir, Paul A. Raschky},
  pdfkeywords={chatgpt, AI, productivity},
  hidelinks,
  pdfcreator={LaTeX via pandoc}}
\usepackage[left=3cm,right=3cm,top=2cm,bottom=2cm]{geometry}
\usepackage{longtable,booktabs}
\usepackage{etoolbox}
\makeatletter
\patchcmd\longtable{\par}{\if@noskipsec\mbox{}\fi\par}{}{}
\makeatother
\IfFileExists{footnotehyper.sty}{\usepackage{footnotehyper}}{\usepackage{footnote}}
\makesavenoteenv{longtable}
\usepackage{graphicx,grffile}
\makeatletter
\def\maxwidth{\ifdim\Gin@nat@width>\linewidth\linewidth\else\Gin@nat@width\fi}
\def\maxheight{\ifdim\Gin@nat@height>\textheight\textheight\else\Gin@nat@height\fi}
\makeatother
\setkeys{Gin}{width=\maxwidth,height=\maxheight,keepaspectratio}
\makeatletter
\def\fps@figure{htbp}
\makeatother
\usepackage{graphicx} 
\usepackage{amsmath}	
\usepackage{tikz}		
\usepackage{caption}			
\usepackage{lscape}
\usepackage{subfigure}

\usepackage[english]{babel}
\usepackage[document]{ragged2e}
\usepackage{hyphenat}
\usepackage[sc]{mathpazo}
\usepackage[T1]{fontenc}
\usepackage{microtype}
\usepackage{setspace}

\usepackage[defaultlines=2,all]{nowidow}
\usepackage{parskip}
\usepackage{booktabs}
\usepackage{longtable}
\usepackage{array}
\usepackage{multirow}
\usepackage{wrapfig}
\usepackage{float}
\usepackage{colortbl}
\usepackage{pdflscape}
\usepackage{tabu}
\usepackage{threeparttable}
\usepackage{threeparttablex}
\usepackage[normalem]{ulem}
\usepackage{makecell}
\usepackage{xcolor}
\usepackage[]{natbib}
\usepackage{multibib}
\newcites{appx}{Appendix References}

\title{The Unintended Consequences of Censoring Digital Technology -- Evidence from Italy's ChatGPT Ban\footnote{We would like to thank Yashdeep Dahiya for outstanding research assistance. }}
\author{David Kreitmeir\footnote{Department of Economics and SoDa Labs, Monash University. See \href{davidkreitmeir.GitHub.io}{davidkreitmeir.GitHub.io} and \url{david.kreitmeir@monash.edu}} \\  Paul A. Raschky\footnote{Department of Economics and SoDa Labs, Monash University. See \href{https://praschky.GitHub.io/}{praschky.GitHub.io} and \url{paul.raschky@monash.edu}}}
\date{\today { }}

\begin{document}
\maketitle
\begin{onehalfspacing}
\thispagestyle{empty}
\begin{abstract}
\justifying
\noindent We analyse the effects of the ban of ChatGPT, a generative pre-trained transformer chatbot, on individual productivity. We first compile data on the hourly coding output of over 8,000 professional GitHub users in Italy and other European countries to analyse the impact of the ban on individual productivity. Combining the high-frequency data with the sudden announcement of the ban in a difference-in-differences framework, we find that the output of Italian developers decreased by around 50\% in the first two business days after the ban and recovered after that. Applying a synthetic control approach to daily Google search and Tor usage data shows that the ban led to a significant increase in the use of censorship bypassing tools. Our findings show that users swiftly implement strategies to bypass Internet restrictions but this adaptation activity creates short-term disruptions and hampers productivity.

\end{abstract}
\vspace{\fill}
\fontsize{11}{12}
\selectfont

\abovedisplayskip=0pt plus 3pt minus 9pt
\abovedisplayshortskip=0pt plus 3pt
\belowdisplayskip=12pt plus 3pt minus 9pt
\belowdisplayshortskip=7pt plus 3pt minus 4pt
\end{onehalfspacing}

 \pagebreak 

\setcounter{page}{1}
\begin{doublespace}
\justifying

\section{Introduction}\label{sec:introduction}

Unrestricted access to the Internet and digital technologies has become fundamental part of society and is an important channel to enhance productivity in modern economies \citep{brynjolfsson2000,jorgenson2001,stiroh2002,hubbard2003,agrawal2019}. Unfortunately, governments across the globe increasingly restrict access to the Internet and digital technologies. Partial or complete shutdowns of a country’s Internet, the restricted access of particular website or the ban of digital technologies is rampant. While government sanctioned Internet shutdowns and censorship is more prevalent in autocratic regimes, some democratic governments also tend to restrict their citizen’s access to the Internet by blocking access to particular websites or apps, often citing cybersecurity or privacy concerns. These government interventions in the digital realm do not only inhibit people’s access to information and their freedom to express opinions and ideas, but they also result in direct economic costs. Business not only rely on the Internet for communication purposes, but increasingly use web and cloud-based technologies as an input factor in their production function. A growing number of AI technologies are hosted in the cloud (i.e. OpenAi) and restricting access to the Internet also bars business from using it.

This paper provides an empirical analysis of the effects of Internet censorship on productivity. In particular, we exploit the sudden announcement of a ban of ChatGPT, a, in Italy and analyse its short-run effects of productivity of GitHub users. We find that the ban of ChatGPT decreased output of Italian GitHub users by around 50\% in the first two business days after the initiation of the ban. Output levels returned to normal levels after that. We also interpret these findings as first evidence that ChatGPT is already actively used for high-skilled tasks in the economy. We further show that Internet users in Italy adapted fairly quickly to the legislation by an increased use of VPN and Tor tools to circumvent the ban. It is likely that the short-term drop in output is a result of users' adaptation behaviour and their time spent to circumvent the ban rather than on productive activities. Overall, our findings reveal that potentially well intended restriction of Internet access, for the purpose of protecting user data and privacy, has unintended, negative consequences on productivity that are short-lived.  

Our paper speaks to the large literature in economics and political science has emphasised on the effects of Internet censorship on various societal outcomes as well as individual reaction to censorship. \citet{roberts2018} documents how the Chinese Firewall adversely affects Internet speed in China while \cite{hobbs2018} shows how Chinese users react to the sudden ban of Instagram by an increased used of  circumvention tools. \citep{chen2019} conduct a field experiment where they randomly provide Chinese Internet users with access to censorship circumvention tools. They find that access to uncensored Internet in itself does not increase the demand for less censored online information. However, in combination with a treatment that encourages the use of uncensored Internet, and the associated consumption of less censored media leads to a change in the level of information and political beliefs. 
In some cases, the blocking of particular websites is also used to counter foreign propaganda and misinformation, such as the ban of the Russian social media platform VKontakte in Ukraine \citep{golovchenko2022}. We complement this literature by focusing on censorship in a Western, democratic country and evaluating its effects on individual productivity of knowledge workers.

We also contribute to the growing economic literature on the effects of artificial intelligence (AI) technology on productivity \citep{brynjolfsson2017,agrawal2019}. Similar to  to time savings through computerisation \citep{autor2003}, AI technology has a large potential to impact more complex and creative tasks. \citet{brynjolfsson2017} argue that the full effects of AI on productivity are difficult to measure in aggregate, national statistics. Analysis at the micro level could provide first insights.  Experimental evidence by \citet{noy23} is among the first to show how ChatGPT can increase output of knowledge workers. They conduct an incentivised, online experiment where participants are asked to perform professional writing tasks. They randomly ask half of the participants to sign up for ChatGPT and explain to them how it can be used to complement their efforts for the latter set of writing tasks. They find that the participants in the treatment group take less time to finish their writing task, the quality of the written output is higher and use of ChatGPT also increases the job satisfaction of the participants. 

\citep{eloundou2023} examine the potential consequences of Generative Pre-trained Transformers (GPTs) on the labor market in the United States, with a focus on the amplified abilities that arise from LLM-powered software as opposed to LLMs alone. Their results indicate that the introduction of LLMs could impact at least 10\% of work tasks for about 80\% of the US workforce, with roughly 19\% of workers potentially experiencing at least a 50\% impact. These effects apply across all wage levels, with higher-paying jobs potentially facing greater exposure to LLM-powered software. Access to an LLM could result in a significant decrease in the time required to complete approximately 15\% of all worker tasks in the US while maintaining the same level of quality. This percentage increases to between 47 and 56\% when incorporating software and tools developed on top of LLMs, suggesting that LLM-powered software could have substantial economic implications. 

 Our study complements existing work by using a large set of secondary data on the output of software and code developers.  Recent studies by \citet{kashefi2023} \citet{sobania2023}  have already highlighted the potential of Chatgpt to complement humans in programming in different programming languages. Among other things, ChatGPT is used for debugging and improving written codes by users, for completing missed parts of numerical codes, rewriting available codes in other programming languages, and for parallelizing serial codes. 

Our paper also relates to the large literature political \citep{acemoglu_2006} religious \citep{benabou2021,cosgel2012} or misaligned incentives within a sector or firm \citep{desmet2014,atkin2017} lead to resistance against the adoption and diffusion of new technological innovations; by workers whose share of income was reduced due the the arrival of new labour-saving technologies \citep{caprettini2020}. 

The paper is organised as follows: Section 2 provides some background information about ChatGPT and the Ban of the technology in Italy; Section 3 describes the data; Section 4 and 5 present empirical results on the effect of the ban on productivity and and adapation behaviour, respectively; Section 5 concludes.

\section{ ChatGPT and its Ban Italy}

ChatGPT, is a large language models (LLM), which was created by US start-up OpenAI, has been used by millions of people since it launched in November 2022. It is a large-scale artificial intelligence (AI) language model that uses a transformer-based neural network to process natural language. ChatGPT has been trained on a vast corpus of online text data from the Internet as it was in 2021. During the training process, the model learned to identify patterns and relationships between words, phrases, and sentences, enabling it to generate text.\footnote{Such as this very paragraph.} 

ChatGPT is accessible via a public website (chatgpt.openai.com) or an API and almost everyone\footnote{Prior to Italy, ChatGpt was already banned in countries like China, Russia or North Korea} can sign up for a free account. The interface is designed like a chat environment where the user writes ``prompts'' and ChatGPT answers. Interactions can range from casual chats, to search like queries, but also more complex interactions such as creative writing of a text based on a prompt, the creation of recipes. ChatGPT also has the capabilities to write code in multiple programming languages based on a simply prompt.

On April 1st 2023, the Italian data protection authority (Garante per la protezione dei dati personali) has blocked the use of the ChatGPT chatbot, citing privacy concerns. The Italian watchdog has announced an investigation into OpenAI's compliance with General Data Protection Regulation (GDPR). In particular authority has said that there was no legal basis to justify the mass collection and storage of personal data for the purpose of training algorithms underlying the operation of the platform.\footnote{Shiona McCallum, ``ChatGPT banned in Italy over privacy concerns'', BBC 01/04/2023, https://www.bbc.com/news/technology-65139406}  

A priori it is unclear how the ban might affect individual productivity. As shown by recent papers, ChatGPT can enhance productivity of coding \citep{kashefi2023,sobania2023} and professional writing tasks \citep{noy23}. Therefore, the sudden ban can have negative effects on individual productivity. However, ChatGPT is still a fairly young technology and there is no reliable data available on its diffusion within the economy. The media hype around the technology might paint a misleading picture of the  adoption by companies and individual knowledge workers. As such, the ban might not have any systematic impact on individual productivity. It is also possible, that 
\section{Data}

\subsection{GitHub Data}

We access individual, real-time activity data from GitHub users in Italy (treatment), Austria and France (control) in the weeks prior and immediately after the ChatGPT ban in Italy, March 27th -- April 11th 2023.

GitHub is the world’s largest online code hosting platform, used for storing and jointly working on coding projects, so called repositories. All modifications to the GitHub repository are automatically timestamped and stored, and GitHub permits the tracking of any iterations to specific files and lines of code. The history of iterations and actions is available for anybody with access to a repository to examine and download. Every action taken by a team member is automatically recorded, keeping details about the kind and substance of the modification, the files and code lines affected, and the date those changes were performed. With GitHub's history of developing open-source software, a significant portion of the repositories are not access restricted, meaning that the information about projects' activity is available to everyone. As such, public GitHub repositories provide a direct, real-time measurement of labor activity of software and code developers for millions of
 software and code developers worldwide \citep{mcdermott21}.

GitHub data has already been used in empirical research to study changes in software developers' productivity during the onset of COVID19 \citep{forsgren2021}, how COVID19  changed the daily and weekly patterns of labour allocation of individuals \citep{mcdermott21}; the impact of the effects of working from home on individuals' productivity \citep{shen2023}, the effect of air pollution on individual output \citep{holub2023}, or analyse the relationship between social links and the likelihood of joining professional software development teams\citep{casalnuovo2015}.

Our preferred measure for GitHub users' output are \textit{Release} events. A release is a deployable version of a software project that make it available for a wider audience and easy for users to download and install the software. It typically includes a set of release notes, documentation, and binaries or source code.\footnote{See https://docs.GitHub.com/en/repositories/releasing-projects-on-GitHub/about-releases}. On an hourly and daily basis, releases are relatively rare events and we create an indicator variable that switches to one if the user has recorded at least one release event during the respective time period (hour or day).

We also use the (log of) number of total GitHub events by a user, $Events$, the sum of Push and Pull events, \textit{PushPull} and an aggregate \textit{Output} measure based on  \citet{holub2023}, which is the sum of Push, Pull Requests, Pull Request Comments, Commit Comments, Create, and Issues. In addition, we use information about the users organisational affiliation to distinguish between GitHub users who have entered an organisational affiliation (i.e. company, research organisation) and those without an organisational affiliation.

We access data on individual GitHub user's activity from and GitHub Archive which is updated daily and contains all public event data. GitHub Archive data is hosted on   Google’s BigQuery warehouse system and can be accessed with a query on google's cloud infrastructure. GitHub user information was downloaded using the GitHub GraphQL API \footnote{The python scripts that was written to access the GitHub user information for the respective countries is available at \url{https://GitHub.com/sodalabsio/GitHub\_scrape}}. Both datasets are then joined via the unique GitHub user login.

Descriptive statistics and a balance test for the GitHub data is presented in Table \ref{tab:descriptive} in the appendix.

\subsection{Google Trends and Tor}

To examine how Italian ChatGPT users adapt after the ban, we collect daily data from two sources \textit{Google Trends} and  \textit{Tor Metrics} for all 25 countries in the European Union. Google trends data has already been widely used in economic research as a predictor of human behaviour economic phenomena \citep{choi2012}. For example \citet{bohme2020} have used Google trends data migration-related google search terms to predict international migration while \citet{ginsberg2009} have used this data to predict influenza outbreaks.

We focus on the period after the release of ChatGPT-4 on 13 March 2023 until the end of the work week after the ban on 7 April 2023. Since we are interested in the effect of the ban on output, observations on weekends are dropped from the sample. Figure \ref{fig:panelView} in the Appendix provides a graphic illustration of the final panel structure.

We construct three outcome measures. First, we obtain the number of \textit{Google searches} on the topic of ``Virtual Private Networks''. \textit{Google} normalizes the number of searches by the total number of searches in the selected country and period and our outcome measure, therefore, is bounded between 0 and 1. Second, we download information on the number of users of \textit{Tor}, an open-source software for enabling anonymous communication. In particular, we collect the number of users of ``standard'' \textit{Tor relays}, and of \textit{Tor bridge} relays. The latter are not listed publicly and, therefore, more difficult to identify for firewalls but can slow down the connection.\footnote{For more information on bridges vs. ``standard'' relays please refer to the official \textit{Tor} documentation at: \url{https://tb-manual.torproject.org/bridges/}.} We apply a log transformation to both user numbers.

\section{The Effect of the Ban on Productivity}

To analyse the effect of the Italian ChatGPT ban on output by GitHub users, we estimate variants of the following Difference-in-Difference (DID) event study model: 
\begin{align}
    Y_{it} =& \alpha_{i} + \lambda_{t} + \sum_{\tau = -3}^{-1} \beta_{\tau}  D_{it}^{\tau} + \sum_{\tau = 0}^{4} \beta_{\tau} D_{it}^{\tau} + \epsilon_{it}, \label{eq:output}
\end{align}

where $Y_{it}$ is an indicator variable that equaling to one if user $i$ made a $Release$ on GitHub on day $t$ and zero otherwise. $D$ is a dummy variable, switching to one for observation’s in the treatment group is in period $\tau$ and zero otherwise. $\alpha_{i}$ is a vector of user specific fixed effects, $\lambda_{t}$ are day (date) fixed effects, $\beta$ are the parameters of interest and $\epsilon_{it}$ is the error term. Standard errors are clustered at the GitHub user-level.

\begin{figure}[!h]
\centering 
\caption{The Effect of the ChatGpt ban on GitHub Releases}
\label{fig:eventstudy}
\includegraphics[width=12cm]{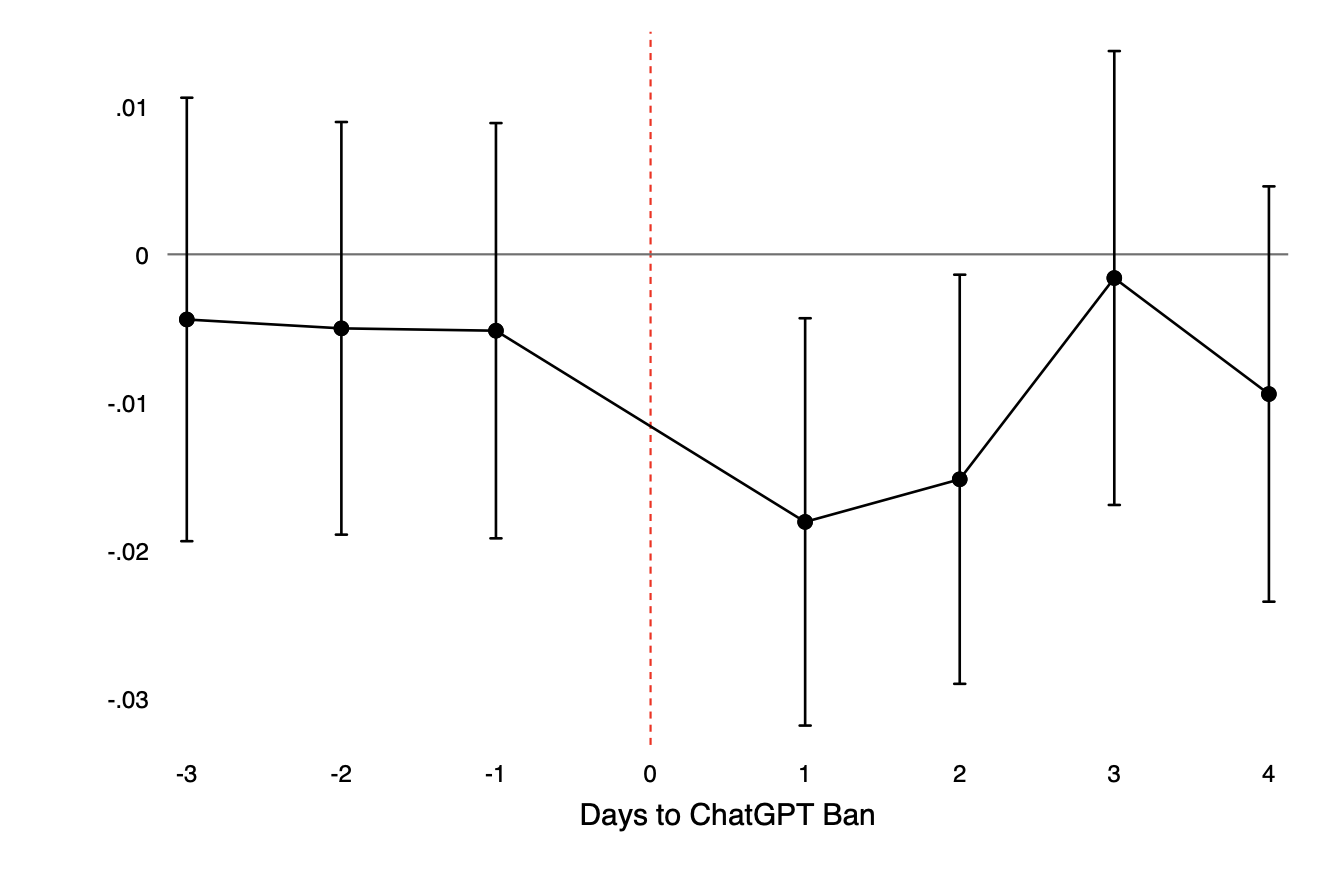}
\captionsetup{justification=justified, singlelinecheck=off} 
\caption*{\small \textit{Notes:} Black dots represent the coefficients from estimating the difference-in-Difference (DID) event study model from equation \ref{eq:output}. 95\% confidence intervals using robust standard errors clustered on the GitHub user-level are depicted. The horizontal axis label denotes the days before and after the ChatGPT ban was implemented. The last day prior to the ban, 31/03/23 is the left out period. The weekend days (01/02.04) after the ban was implemented are not included.}
\end{figure}

The results are presented in Figure \ref{fig:eventstudy} showing a statistically significant drop in release likelihood in the two first business days after the ChatGPT ban was initiated. 

We then proceed and estimate a difference-in-differences specification that combines the $Pre$ and $Post$ period. The results are presented in Table \ref{tab:baseline}. In the first specification in panel $A$, we specify the pre period as the last Monday and Tuesday (March 27th and 28th 2023) prior to the ban and the post period as the first Monday and Tuesday (April 3rd and 4th 2023) after the implementation of the ban. Columns 1--4 present estimates for the sample of users with organisational affiliation, while columns 5--8 are the estimates for users without affiliation. Starting with the sample of GitHub users with an organisational affiliation, we find no systematic effect on the ChatGPT ban on the total number of GitHub events. However, we find a a statistically significant and negative effect on the likelihood of $Release$ events. Considering the mean of $Release$ in the sample of 0.024, the point translates into an approx. 50\% reduction in the likelihood that a developer will have a release event recorded on GitHub in the first two business days after the ban. The estimated coefficients for $Output$ and $PushPull$ also have a negative sign but are not statistically significant. We do not find any systematic effects of the ban on activity by GitHub users without an organisational affiliation. Panel $B$ contains results where the post period consists of the next three business days after the ban, Thursday, Friday (April 5th and 6th)\footnote{We excluded April 7th because this was Good Friday} as well as Monday (April 11th). None of the estimated coefficients of $Treat \times Post$ is statistically significant and the coefficients also have a lower magnitude compared to the results in panel $A$. This suggests that the disruptions caused by the ban was relatively short-lived and Italian GitHub users quickly recovered their pre-ban levels of productivity.

 \begin{table}[!h]
\caption{\label{tab:baseline}The Effect of the ChatGPT Ban on GitHub Output }
\centering
\begin{threeparttable}
\begin{tabular}[t]{lcccccccc}
\toprule
&\multicolumn{4}{c}{ \textbf{Users with Affiliation}} &\multicolumn{4}{c}{\textbf{Users w/o Affiliation}}\\
&Events &  Release   &     Output  & PushPull & Events &  Release   &     Output  & PushPull     \\ 
\cline{2-9} 
 & (1) & (2) & (3) & (4) & (5) & (6) & (7) & (8)   \\
\midrule
&\multicolumn{8}{c}{\textit{A: Pre 27/28.03 - Post 03/04.04}} \\ 
\cline{2-9} 

Treat$\times$Post&       0.115   &      -0.012** &      -0.278   &      -0.227   &       0.005   &      -0.003   &       0.188   &       0.192   \\
            &     (0.109)   &     (0.005)   &     (0.295)   &     (0.207)   &     (0.120)   &     (0.005)   &     (0.221)   &     (0.189)   \\
Post  &       0.052   &       0.002   &       0.453***&       0.323***&      -0.013   &       0.002   &       0.040   &       0.051   \\
            &     (0.053)   &     (0.002)   &     (0.123)   &     (0.094)   &     (0.060)   &     (0.002)   &     (0.119)   &     (0.101)   \\
\midrule
 
N           &   18332  &   18332    &   18332   &   18332   & 14496   &   14496   &   14496   &   14496   \\
\midrule
&\multicolumn{8}{c}{\textit{B: Pre 27/28/29.03 - Post 05/06/11.04}} \\ 
\cline{2-9}

Treat$\times$Post&       0.048   &      -0.003   &       0.135   &       0.073   &       0.003   &       0.000   &       0.067   &       0.058   \\
            &     (0.090)   &     (0.004)   &     (0.223)   &     (0.163)   &     (0.101)   &     (0.004)   &     (0.175)   &     (0.142)   \\
Post &      -0.217***&      -0.002   &      -0.037   &      -0.001   &      -0.325***&      -0.000   &      -0.112   &      -0.069   \\
            &     (0.042)   &     (0.002)   &     (0.097)   &     (0.074)   &     (0.050)   &     (0.002)   &     (0.095)   &     (0.078)   \\\midrule

N           &   31983   &   31983   &   31983   &   31983   &   25368   &   25368  &   25368  &   25368  \\
 
\bottomrule
\end{tabular}
\begin{tablenotes}[para]
\item \textit{Notes:} 
\item All specifications include user fixed effects. Robust standard errors are clustered on the user-level in parentheses: *p<0.1, ** p<0.5, *** p<0.01.
\end{tablenotes}
\end{threeparttable}
\end{table}

Table \ref{tab:heterogeneity} presents the results of a heterogeneity analysis using the sample of users with an organisational affiliation. The results reveal that the adverse effects of the ban on productivity are mainly driven by users who created their profile prior to 2016 and users with 15 followers or less. While the latter could be an indicator for less skilled users, the former results suggests that older users are more impacted by the ban.

 \begin{table}[!h]
\caption{\label{tab:heterogeneity}Heterogeneity Analysis }
\centering
\begin{threeparttable}
\begin{tabular}[t]{lcccccccc}
\toprule
&Events &  Release   &     Output  & PushPull & Events &  Release   &     Output  & PushPull     \\ 
\cline{2-9} 
 & (1) & (2) & (3) & (4) & (5) & (6) & (7) & (8)   \\
\midrule 
&\multicolumn{4}{c}{\textit{Profile created prior 2016}} &\multicolumn{4}{c}{\textit{Profile created after 2016}} \\ 
\cline{2-9}  

Treat$\times$Post&       0.068   &      -0.015** &      -0.131   &      -0.082   &       0.157   &      -0.008   &      -0.457   &      -0.404   \\
            &     (0.144)   &     (0.007)   &     (0.461)   &     (0.293)   &     (0.160)   &     (0.006)   &     (0.326)   &     (0.273)   \\
Post  &       0.039   &       0.001   &       0.291*  &       0.214   &       0.052   &       0.001   &       0.596***&       0.420***\\
            &     (0.068)   &     (0.004)   &     (0.170)   &     (0.130)   &     (0.080)   &     (0.003)   &     (0.165)   &     (0.125)   \\ 
\midrule
N           &   10192   &   10192   &   10192   &   10192   &    8728   &    8728   &    8728  &    8728    \\

\midrule
&\multicolumn{4}{c}{\textit{GitHub Followers $\leq$ 15}} &\multicolumn{4}{c}{\textit{GitHub Followers $>$ 15}} \\ 
\cline{2-9} 
Treat$\times$Post&       0.092   &      -0.011** &      -0.879*  &      -0.601*  &       0.131   &      -0.011   &       0.359   &       0.158   \\
            &     (0.158)   &     (0.005)   &     (0.461)   &     (0.322)   &     (0.143)   &     (0.008)   &     (0.321)   &     (0.228)   \\
Post  &       0.052   &      -0.004   &       0.619***&       0.467***&       0.038   &       0.007*  &       0.243   &       0.150   \\
            &     (0.076)   &     (0.003)   &     (0.158)   &     (0.129)   &     (0.070)   &     (0.004)   &     (0.178)   &     (0.128)   \\
\midrule
N           &    9532   &    9532   &    9532    &    9532   &    9388   &    9388   &    9388    &    9388   \\

\bottomrule
\end{tabular}
\begin{tablenotes}[para]
\item \textit{Notes:} 
\item Only users with an organisational affiliation. Period Pre 27/28.03 - Post 03/04.04. Specifications include user fixed effects. Robust standard errors are clustered on the user-level in parentheses: *p<0.1, ** p<0.5, *** p<0.01.
\end{tablenotes}
\end{threeparttable}
\end{table}

In Table \ref{tab:placebo} we report the result of a placebo test. While ChatGPT can have effects on GitHub events that require some productive activities, the writing and debugging of code is a pre-requisite for a stable $Release$, for other GitHub activities such as $Gollum$, $Member$ or $Public$ there is no use for ChatGPT. For those type of activities we would not expect an systematic impact of the ban of ChatGPT. The result in Table \ref{tab:placebo} show that the implementation has basically a null effect on those activities. 

In Figure \ref{fig:dailyrelease} we compare the average, hourly release likelihood between Italian GitHub users and GitHub users from the control countries over the period March 27th to April 6th in UTC time\footnote{Austria, Italy and France time zones are UTC +2 hours. 6:00 in Figure \ref{fig:dailyrelease} corresponds to 8:00 local time.}. The first row (panels a--d) contains plots for the Monday to Thursday prior to the implementation of the ban; the second row contains plots for the Monday to Thursday in the first week after the implementation of the ban; and the third row contains a plot for Friday, March 31st when the ban was announced, Saturday April 1st, when the ban started and Sunday, April 1st, which was the first(second) weekend day the ban was in place. The red line are the average, hourly release likelihood values for the Italian users, and the blue one are the ones for the control countries. In most of the weekdays plots (a--i) one can see the intradiurnal patterns of a typical workday. Output is relatively low in the early morning and starts to increase at around 8:00 and then dropping again after 18:00--20:00 local time. On the business days prior to the introduction of the ban (panels a--d), output of Italian and control users was vary similar during the day. There was also no difference in GitHub release patterns on the day the ban was announced, Friday March 31st (panel i) as well as on the following Saturday and Sunday (panels j--k). Output on those two days was also comparatively low. One interpretation is that on those days mainly users who do not rely on ChatGPT for their coding work are active or users would mainly work from home on the weekend where users have already set up (or it is easier to set up) a private VPN or Tor connection to circumvent the ban.
In addition, our results also provide first indicative evidence that ChatGPT, despite being only a few months old and still considered to be in beta version, is already systematically used to support software and coding activities.

\section{Adaption to the ChatGPT Ban}\label{sec:google-tor}

Considering the relatively quick recovery of individual productivity after the implementation of the ChatGPT ban, we now turn our attention towards adaptation behaviour. The simplest way to circumvent the ChatGPT ban is the use of VPN tools or encrypted routing through the Tor network. We conduct our analysis at the country-workday level. To estimate the average treatment effect of the Chat GPT ban on users from Italy, we apply the generalized synthetic control method proposed by \citet{xu2017}. The treatment effect on the treated unit (ATT) is the difference between the actual outcome and its estimated counterfactual. To obtain the counterfactual, a (cross-validated) interactive fixed effects (IFE) model is estimated for the control group data.\footnote{In specific, we apply the EM algorithm proposed by \cite{Gobillon2016} and implemented in the \textbf{R} package \texttt{gsynth} \citep{gsynth2022}, which additionally uses treatment group information for the pre-treatment period, leading to (slightly) more precisely estimated coefficients.} All IFE models incorporate additive unit and time fixed
effects.\footnote{Note that the \textit{Google trends} data is already standardized by country for the selected time period. Hence, the estimated IFE model in this case only incorporates time fixed effects.} To draw inference, we rely on the parametric bootstrap procedure suggested by \cite{xu2017} in settings with a small number of treated units. 

\begin{figure}[!h]
\centering 
\caption{The Effect of the ChatGPT ban on Google Search of VPN}
\label{fig:google-vpn}
\includegraphics[width=\textwidth]{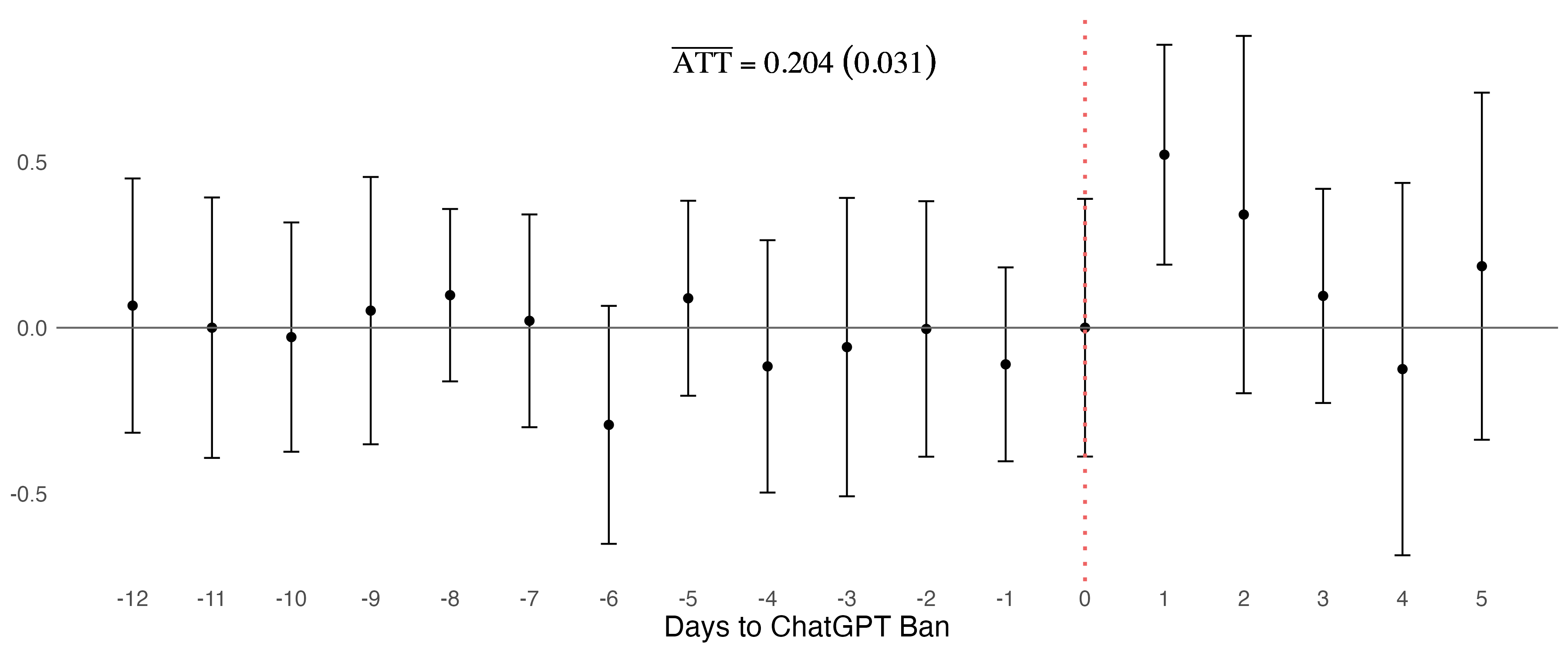}
\captionsetup{justification=justified, singlelinecheck=off} 
\caption*{\small \textit{Notes:} The dynamic treatment effects estimates for the generalized synthetic control method of \cite{xu2017} are depicted. The counterfactual for the treated unit (Italy) is estimated using an interactive fixed effects model. 95\% confidence intervals using the parametric bootstrap procedure proposed by \cite{xu2017} are displayed. Additionally, the average ATT over the workweek after the ChatGPT ban and its $p$-value (in parentheses) are presented.}
\end{figure}

Figure \ref{fig:google-vpn} presents the effect of the ChatGPT ban on the number of \textit{Google queries} on the topic of VPNs. We observe a significant increase in the share of web searches relative to other topics in Italy on the first working day after the ban that slowly vanishes over the next three days. The estimated effect on April 3rd is sizeable: the share of searches on VPNs increases by 52.2 percentage points. On average, the share of queries on VPNs was 20.6 percentage points higher in Italy over the workweek. The observed pattern is consistent with Italian users looking for ways to access ChatGPT even after the ban and succeeding after some initial search costs. Our estimates might, however, present only \textit{stated} preferences. 

\begin{figure}[!h]
\centering 
\caption{The Effect of the ChatGPT ban on Tor Usage}
\label{fig:tor}
\includegraphics[width=\textwidth]{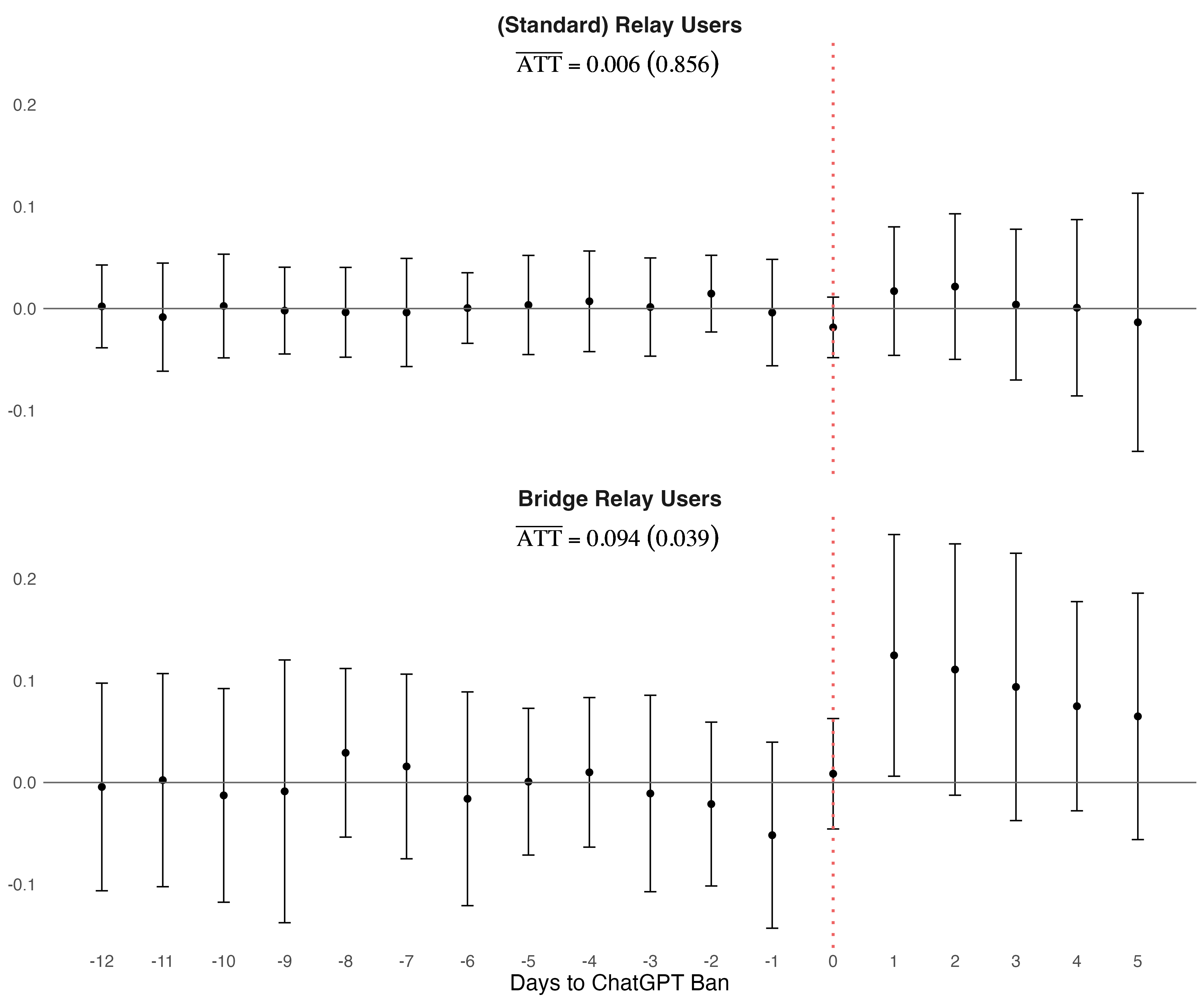}
\captionsetup{justification=justified, singlelinecheck=off} 
\caption*{\small \textit{Notes:}  The dynamic treatment effects estimates for the generalized synthetic control method of \cite{xu2017} are depicted. The top panel presents the average treatment effect on the log number of ``standard'' \textit{Tor relay} users in Italy. The bottom panel presents the ATT for \textit{Tor bridge} relay users. The counterfactual for the treated unit (Italy) is estimated using an interactive fixed effects model. 95\% confidence intervals using the parametric bootstrap procedure proposed by \cite{xu2017} are displayed. Additionally, the average ATT over the workweek after the ChatGPT ban and its $p$-value (in parentheses) are presented.}
\end{figure}

To investigate if the ban actually lead to behavioral changes of Italian users, we look at an alternative outcome: the log number of \textit{Tor} users. The results for \textit{Tor relay} (top panel) and \textit{Tor bridge} users (bottom panel) are presented in Figure \ref{fig:tor}. While the number of \textit{Tor relay} users only shows a minor increase in the days after the ban, the average treatment effect on the number of Italian \textit{Tor bridge} users is positive and significant on the first working day after ban. The usage of \textit{Tor bridges} remains elevated for the entire workweek, with an increase in user numbers--on average--of about 9.4 percentage points. This pattern is in line with users resorting to \textit{bridge} over ``standard'' relays to minimize the chance of being denied access from ChatGPT, since the former are more difficult to identify by firewalls.\footnote{For a discussion on the denial of access to ChatGPT, please see the following OpenAI forum discussion: \url{https://community.openai.com/t/access-denied-error-1020/38758/23}.}

Overall, our findings provide support for the notion that the sizable but short-lived decline in output after the ban is the result of Italian users looking for and successfully finding ways of circumventing the blocked access to ChatGPT.

\pagebreak
\section{Conclusion}

We study the consequences of the ban of ChatGPT, a web-based, generative AI technology, in Italy. We compile high-frequency data on GitHub activity from over 8,000 users in Italy and other European countries to measure individual level output of software and code developers. We show that the sudden ban of ChatGPT decreased output by Italian users of around 50\% in the first two days after the ban. We do not find any effects on output after that. This pattern is likely driven by Italian users' efforts to bypass the ban. Using Google search trend and Tor data in a synthetic control design we find that searches for VPN increases by around 52 percentage points in the days after the ban while the usage of Tor bridges increases by 9.4 percentage points. 

Our findings indicate that government-mandated blocking of digital technology can have adverse effects on the economy. While these measures may be well-intented, they are often ineffective and can lead to short-term disruptions in output. Sudden bans can be easily circumvented with VPN tools, but these adjustment activities simultaneously distort production processes and negatively impact productivity in professions that rely on the banned technology. This can ultimately lead to short-run disruptions in economic output. Overall, our research highlights the need for policymakers to consider the potential economic cost of digital technology bans before imposing them.




\end{doublespace}

\pagebreak

\bibliographystyle{apalike}
\bibliography{bibliography.bib}

\clearpage

\appendix
\renewcommand{\thesection}{\Alph{section}}

\numberwithin{table}{section} 
\numberwithin{figure}{section} 
\numberwithin{equation}{section}

\numberwithin{table}{section} 
\numberwithin{figure}{section} 
\numberwithin{equation}{section}

\setcounter{page}{1} 

\section*{\huge Internet Appendix}

\section{Additional Tables}

 \begin{table}[!h]
\caption{\label{tab:descriptive} Descriptive Statistics }
\centering
\begin{threeparttable}
\begin{tabular}{@{\extracolsep{5pt}}lcccccc}
\\[-1.8ex]\hline \hline \\[-1.8ex]

 & \multicolumn{2}{c}{\textbf{Control}}  & \multicolumn{2}{c}{\textbf{Treatment}}  & &\multicolumn{1}{c}{\textbf{Diff \&}}  \\
\textbf{Variable} & N & \textbf{Mean} & N & \textbf{Mean}  & N & \multicolumn{1}{c}{\textbf{t-test}}\\\hline \\[-1.8ex] 
\# of followers   & 6407    & 60.768    & 1786    & 55.699    & 8193    & 5.069   \\
 &   & (1.19e+05)  &   & (78776.127)  &   &  \\ [1ex]
\# of repositories   & 6407    & 42.568    & 1786    & 39.597    & 8193    & 2.971   \\
 &   & (5616.802)  &   & (3054.544)  &   &  \\ [1ex]
Year GitHub   & 6407    & 2015.956    & 1786    & 2016.476    & 8193    & -0.520***   \\
profile created  &   & (14.956)  &   & (15.454)  &   &  \\ [1ex]
Organisation (Y/N)   & 6407    & 0.563    & 1786    & 0.540    & 8193    & 0.022*   \\
 &   & (0.246)  &   & (0.249)  &   &  \\ [1ex]
Pre Release   & 6407    & 0.020    & 1786    & 0.023    & 8193    & -0.003   \\
 &   & (0.009)  &   & (0.010)  &   &  \\ [1ex]
Pre Events   & 6407    & -1.050    & 1786    & -0.995    & 8193    & -0.056   \\
 &   & (5.358)  &   & (5.232)  &   &  \\ [1ex]
Pre Output   & 6407    & 3.642    & 1786    & 3.475    & 8193    & 0.167   \\
 &   & (27.572)  &   & (27.388)  &   &  \\ [1ex]
Pre PushPull   & 6407    & 2.587    & 1786    & 2.519    & 8193    & 0.067   \\
 &   & (15.012)  &   & (14.806)  &   &  \\ [1ex]

\hline \hline 
\end{tabular}
\begin{tablenotes}[para]
\item \textit{Notes:} Descriptive Statistics. Author's calculations based on GitHub activity data. Cross sectional data.
\end{tablenotes}
\end{threeparttable}
\end{table}

 \begin{table}[!h]
\caption{\label{tab:placebo} Placebo Test }
\centering
\begin{threeparttable}
\begin{tabular}[t]{lcccccc}
\toprule
&Gollum &  Member   &      Public &Gollum &  Member   &      Public    \\ 
\cline{2-7} 
 & (1) & (2) & (3) & (4) & (5) & (6)   \\
\midrule
&\multicolumn{6}{c}{\textit{A: Pre 27/28.03 - Post 03/04.04}} \\ 
\cline{2-7} 
 
Treat$\times$Post&       -0.001   &      -0.001   &      -0.000   &      -0.001   &       0.000   &      -0.001   \\
            &     (0.002)   &     (0.003)   &     (0.003)   &     (0.002)   &     (0.004)   &     (0.004)   \\
Post  &       0.001   &      -0.000   &       0.001   &       0.002*  &       0.000   &      -0.000   \\
            &     (0.001)   &     (0.001)   &     (0.001)   &     (0.001)   &     (0.002)   &     (0.001)   \\\midrule
 
N           &   18332  &   18332       &   18332   & 14496   &   14496   &   14496      \\
 
\bottomrule
\end{tabular}
\begin{tablenotes}[para]
\item \textit{Notes:} LPM estimates. All outcome variables are indicator variables that switch to one if an event of a specific type occurred and zero otherwise. Specifications include user fixed effects. Robust standard errors are clustered on the user-level in parentheses: *p<0.1, ** p<0.5, *** p<0.01.
\end{tablenotes}
\end{threeparttable}
\end{table}

\pagebreak 
\section{Additional Figures}
\begin{landscape}
\begin{figure}
    \centering
       \caption{GitHub Releases - Hourly Data March 27th -- April 6th 2023}
    \subfigure[MON Pre]{\includegraphics[width=0.35\textwidth]{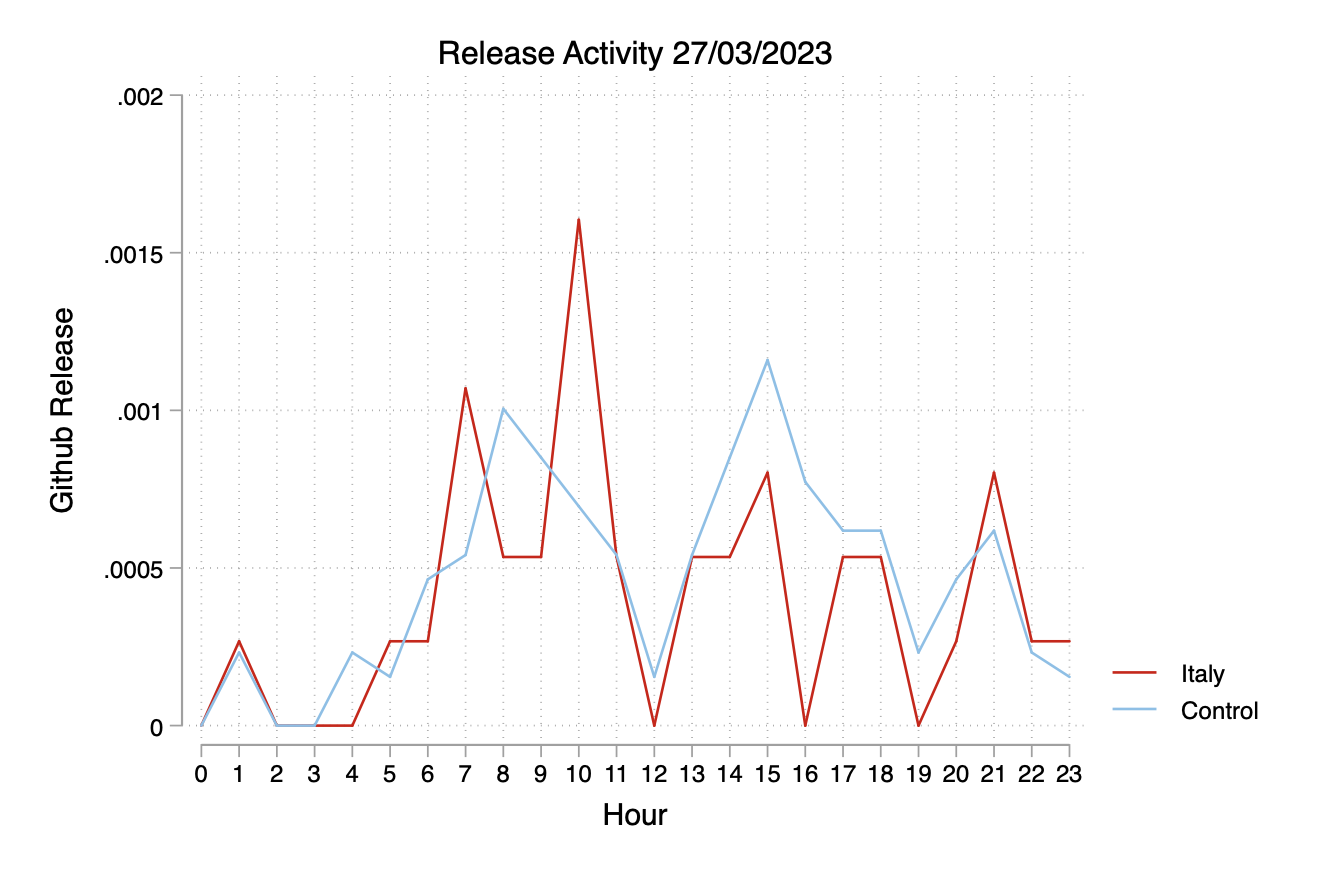}} 
    \subfigure[TUE Pre]{\includegraphics[width=0.35\textwidth]{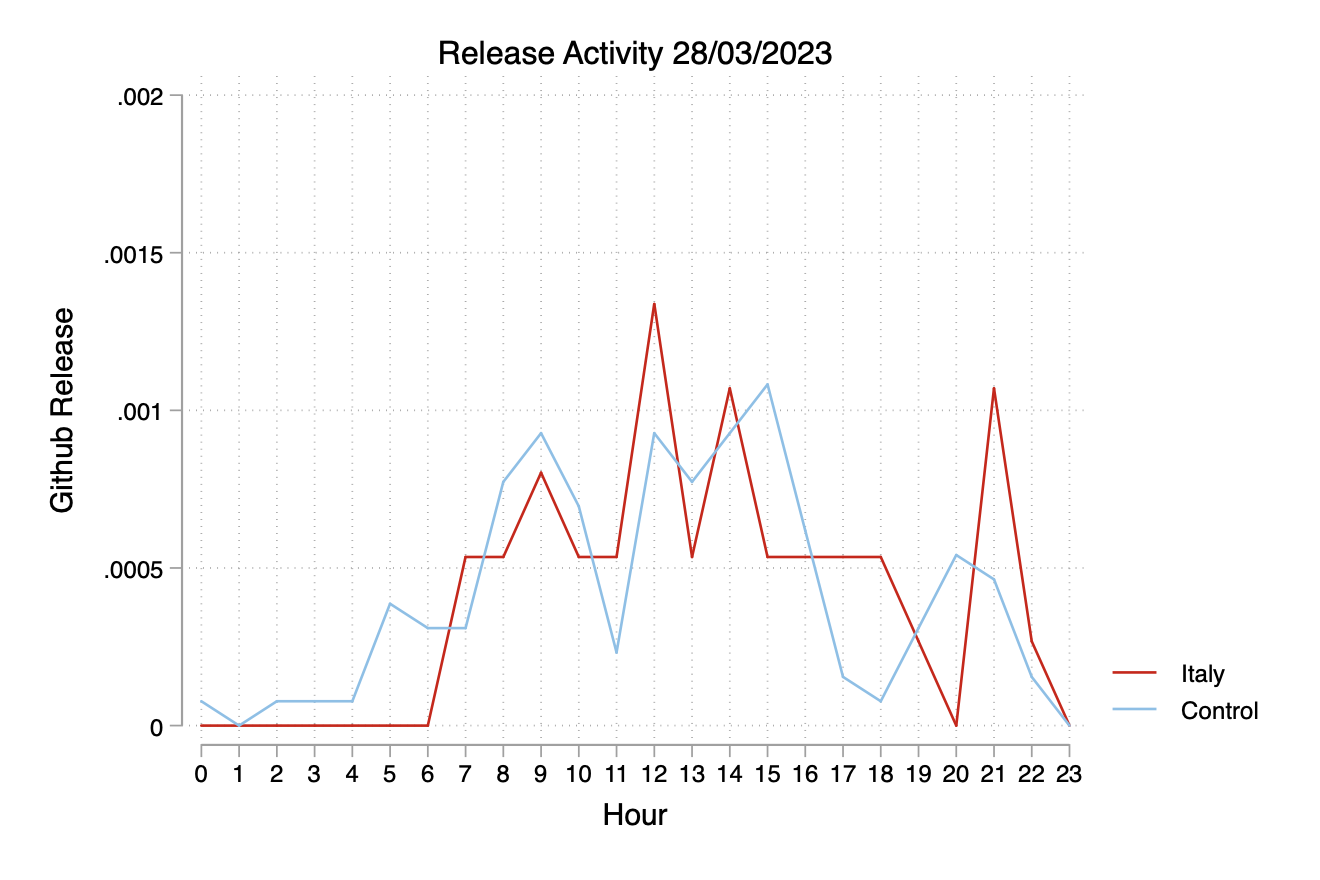}}
    \subfigure[WED Pre]{\includegraphics[width=0.35\textwidth]{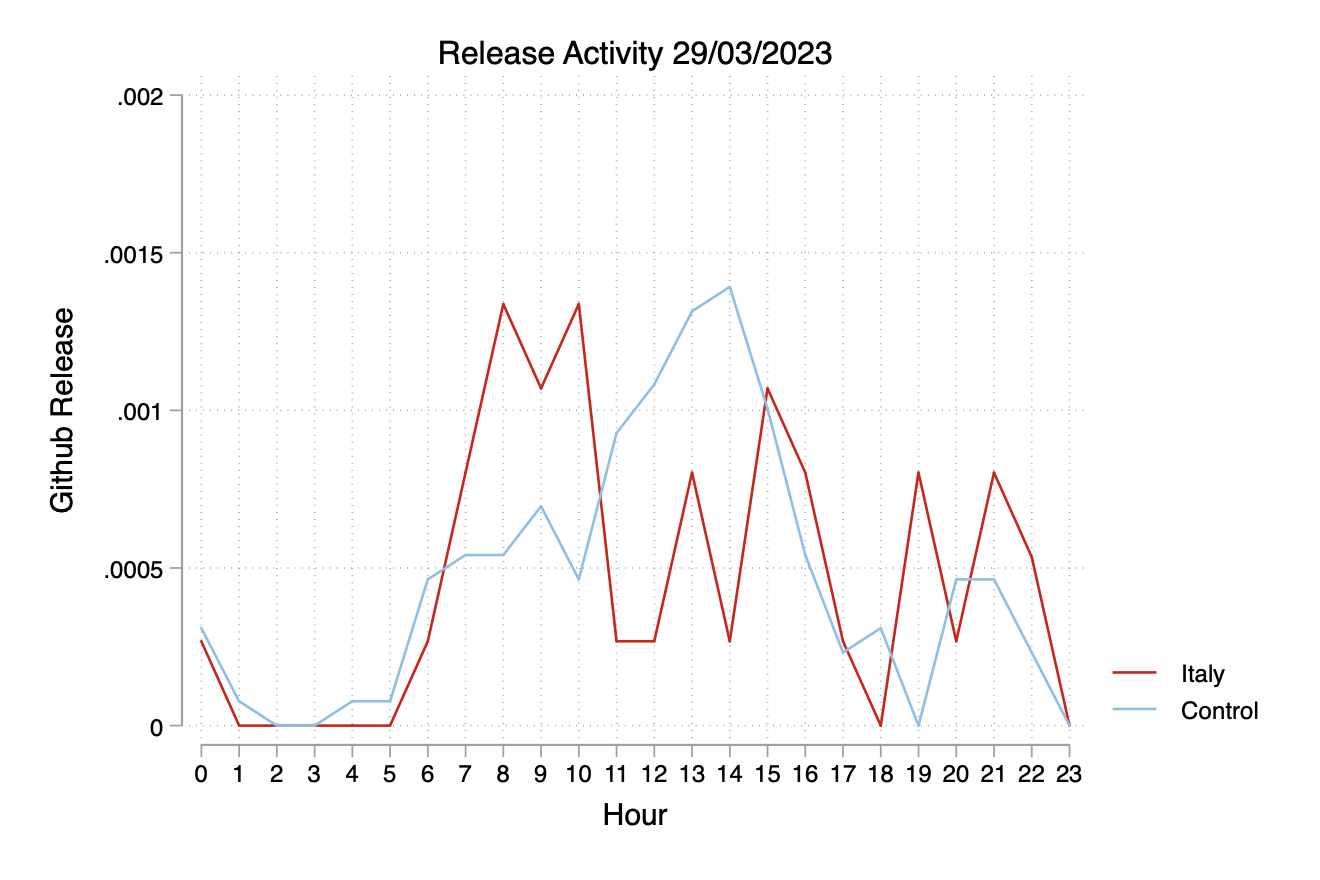}}
    \subfigure[THU Pre]{\includegraphics[width=0.35\textwidth]{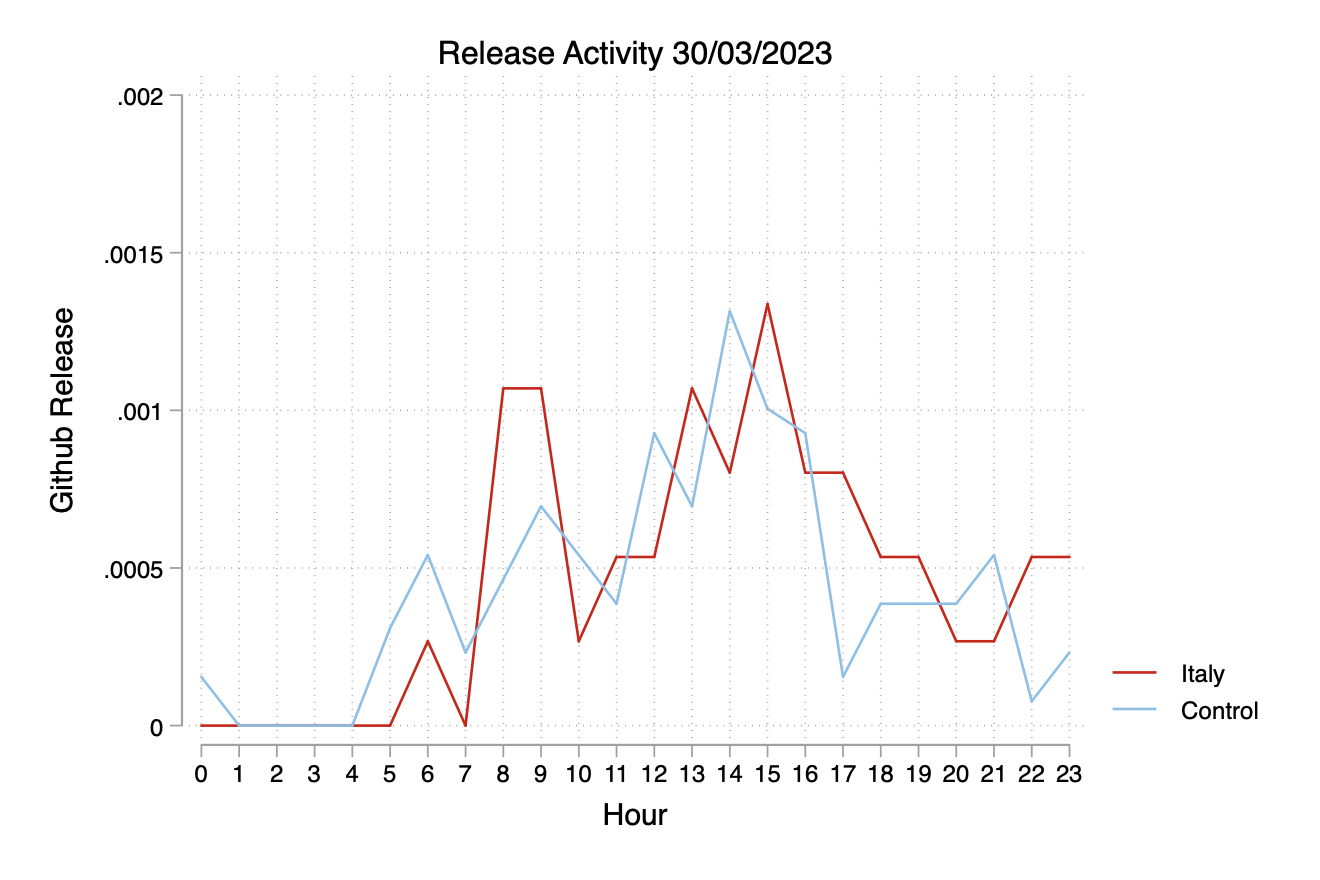}}\\
      \subfigure[MON Post]{\includegraphics[width=0.35\textwidth]{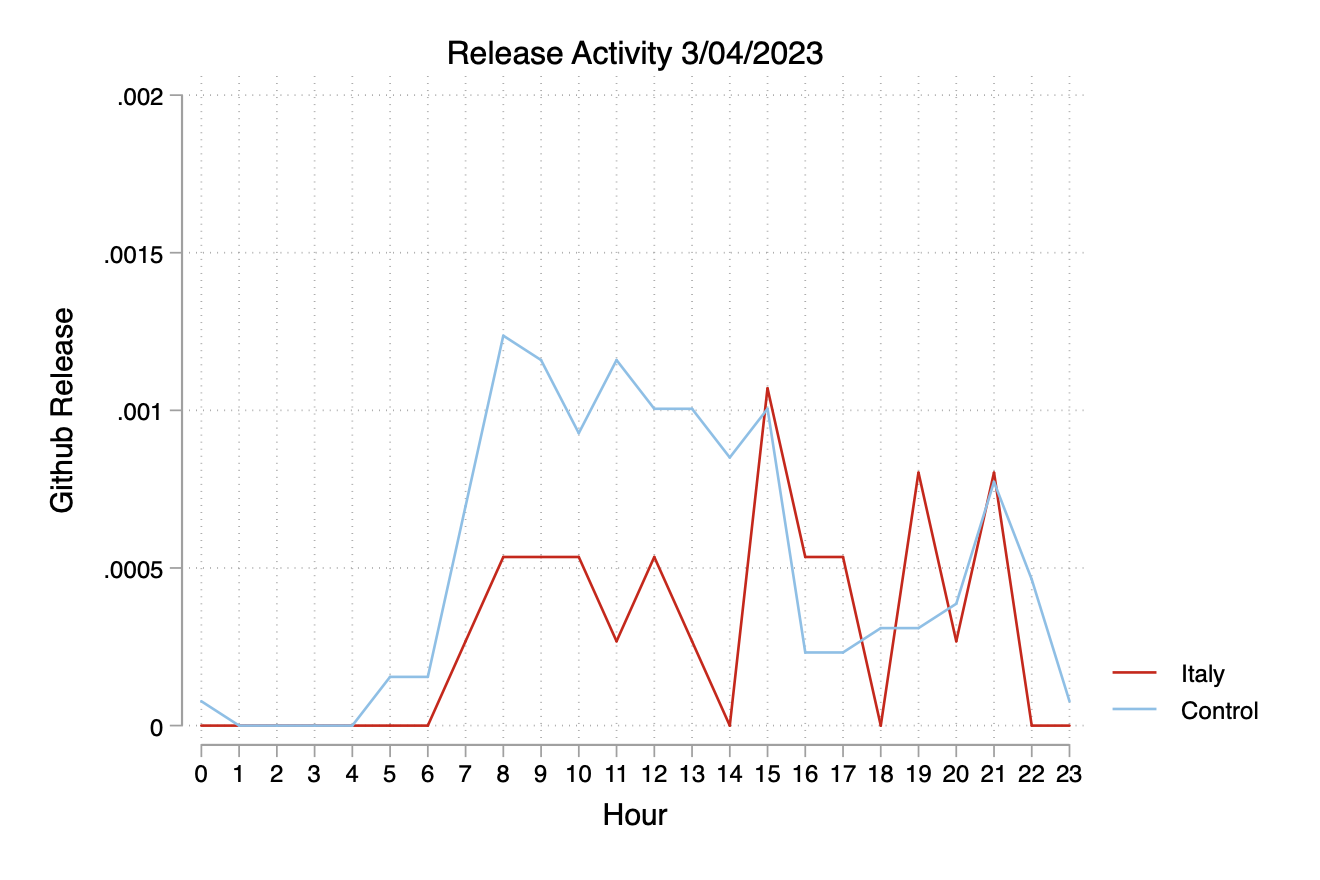}} 
    \subfigure[TUE Post]{\includegraphics[width=0.35\textwidth]{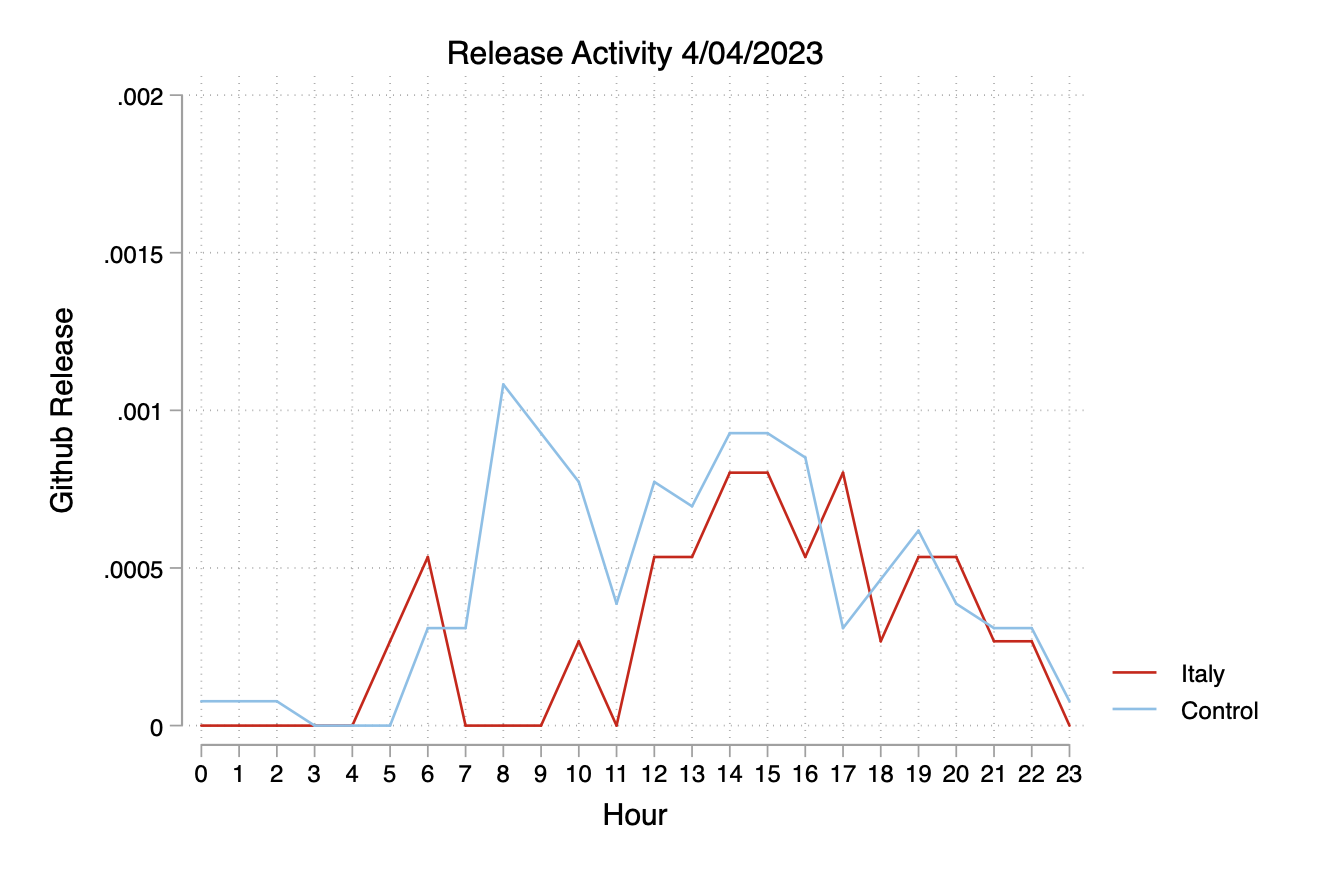}}
    \subfigure[WED Post]{\includegraphics[width=0.35\textwidth]{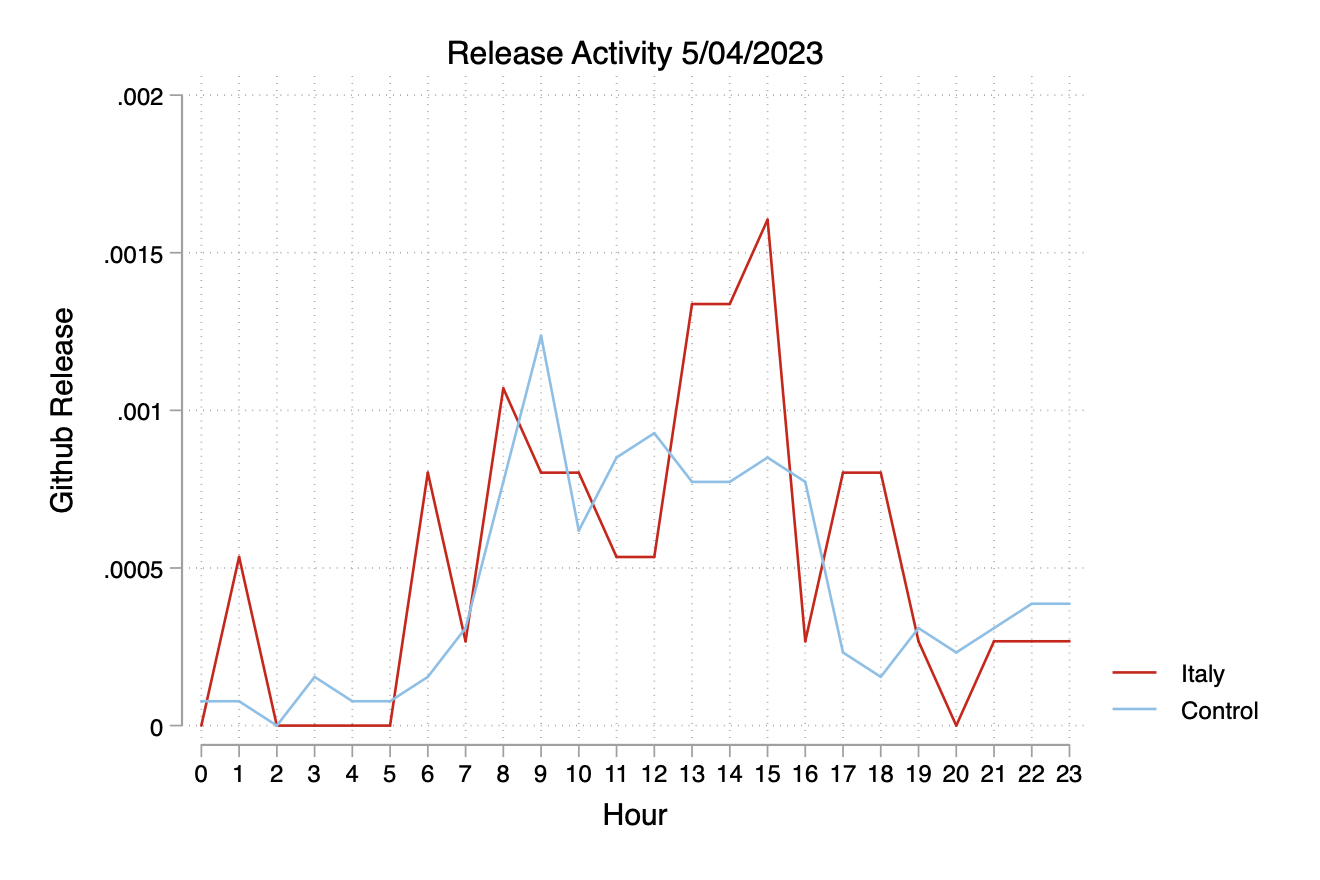}}
    \subfigure[THU Post]{\includegraphics[width=0.35\textwidth]{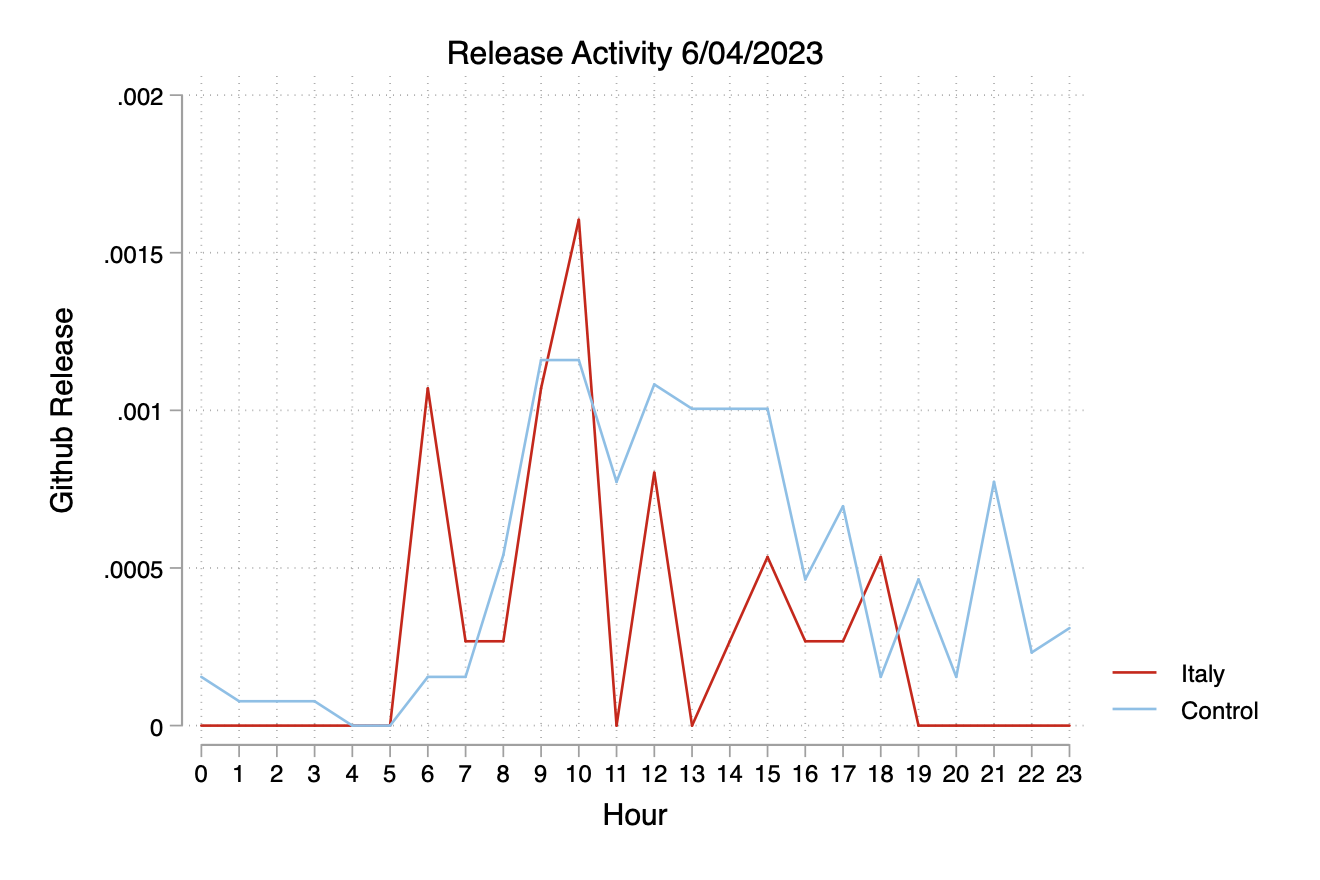}}\\
      \subfigure[FRI Announcement]{\includegraphics[width=0.35\textwidth]{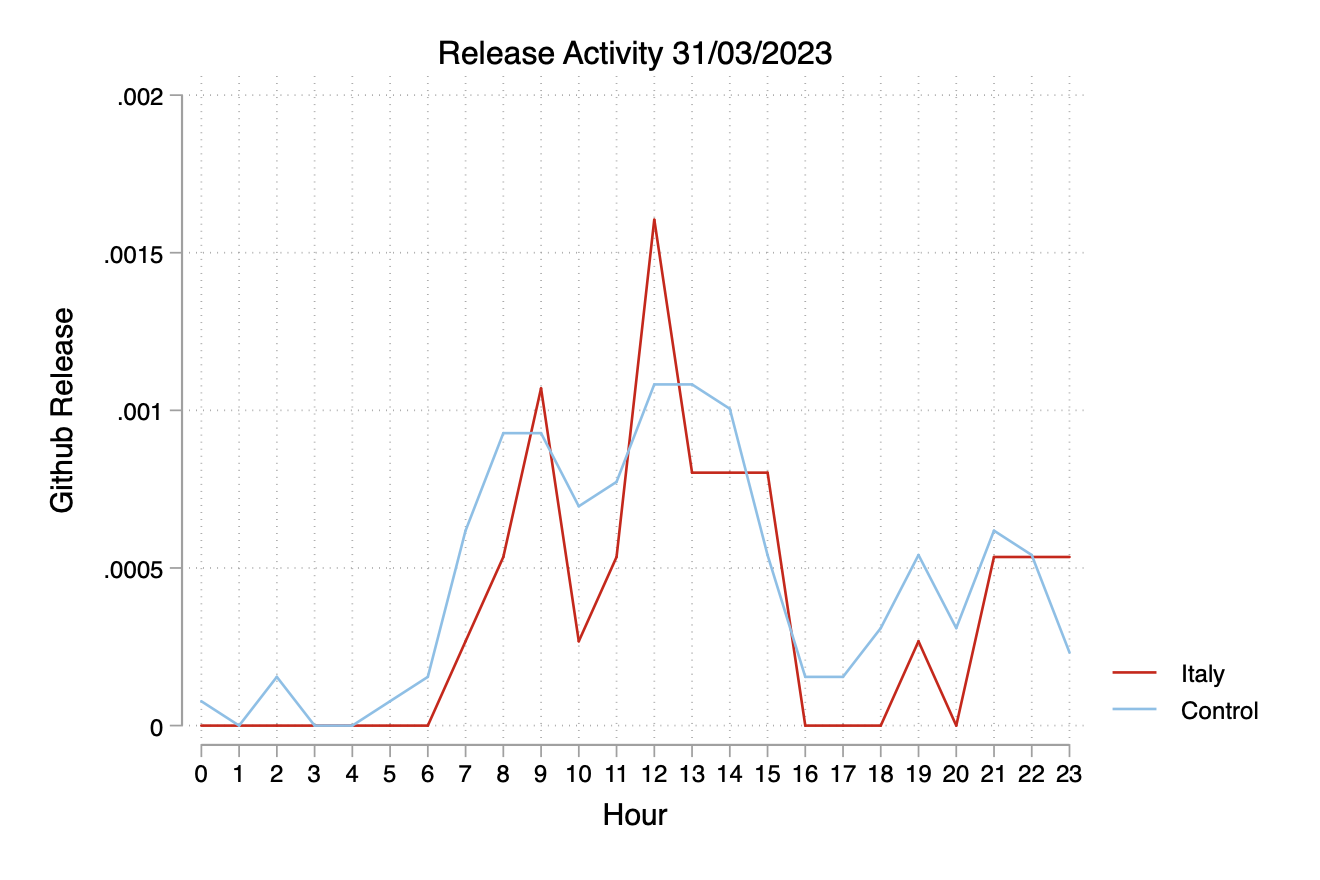}} 
    \subfigure[SAT Ban]{\includegraphics[width=0.35\textwidth]{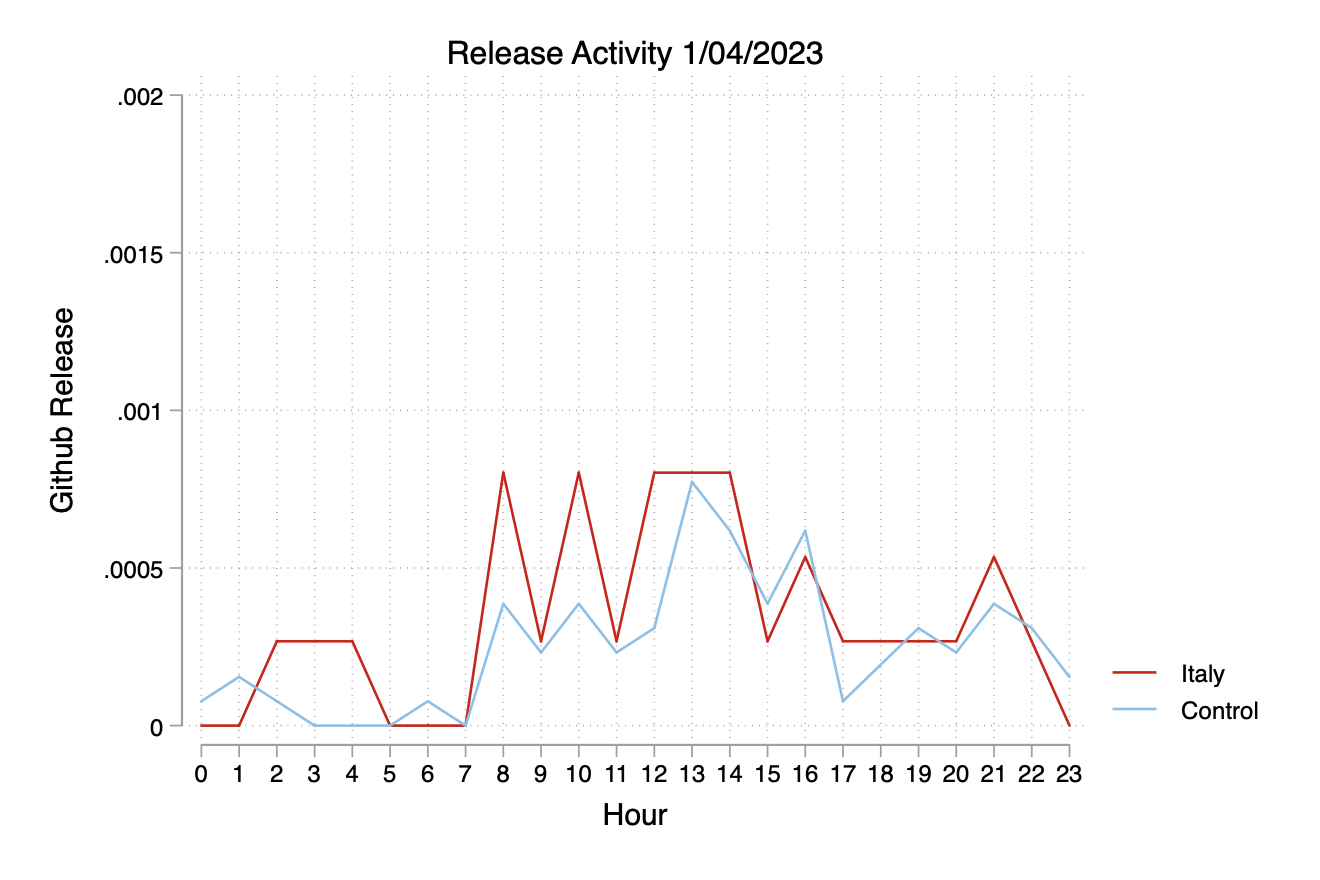}}
    \subfigure[SUN Post]{\includegraphics[width=0.35\textwidth]{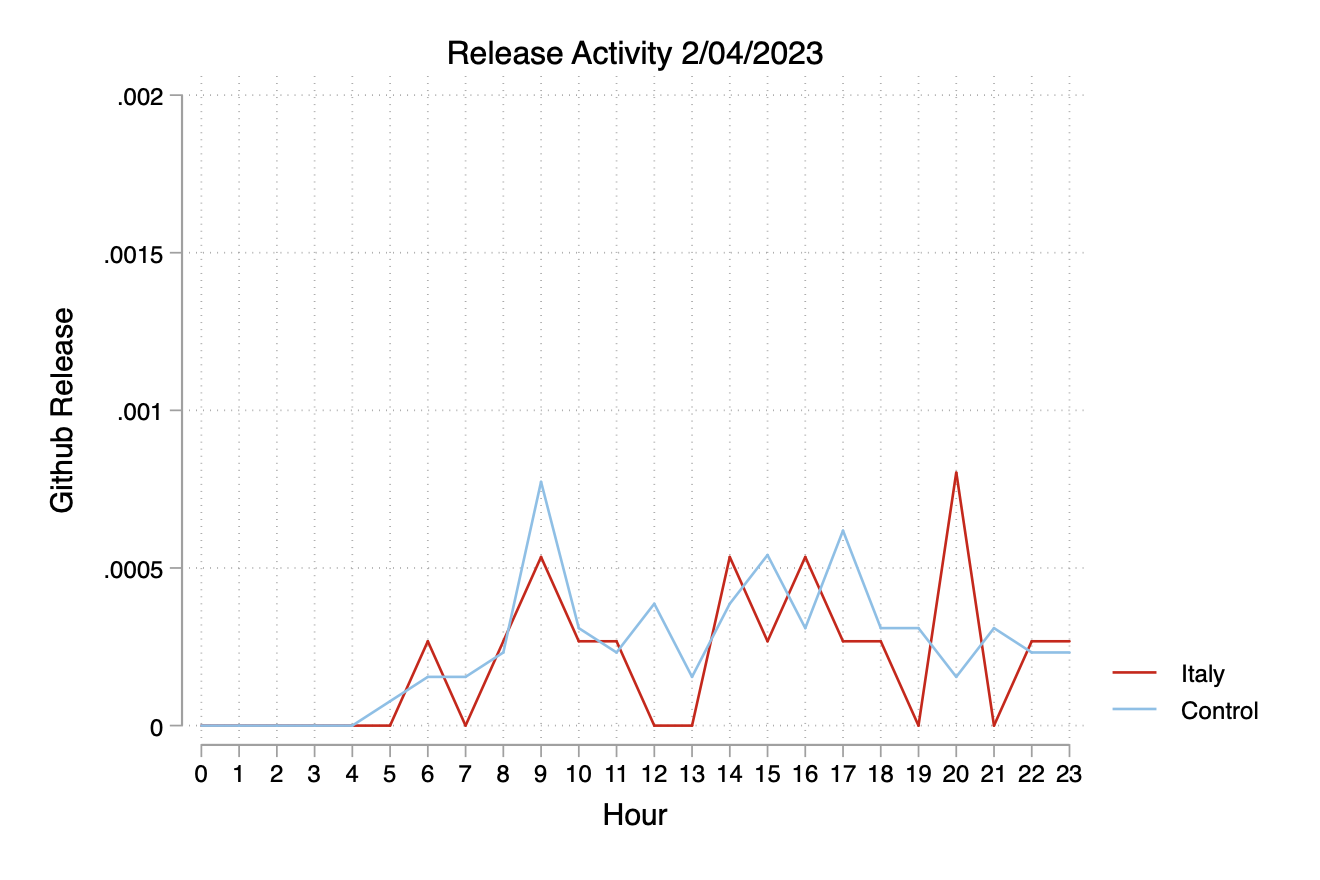}}
 
    \label{fig:dailyrelease}
\end{figure}
\end{landscape}
\clearpage

\begin{figure}[!h]
\centering 
\caption{Panel Structure of Google Trends and Tor Data}
\label{fig:panelView}
\includegraphics[width=\textwidth]{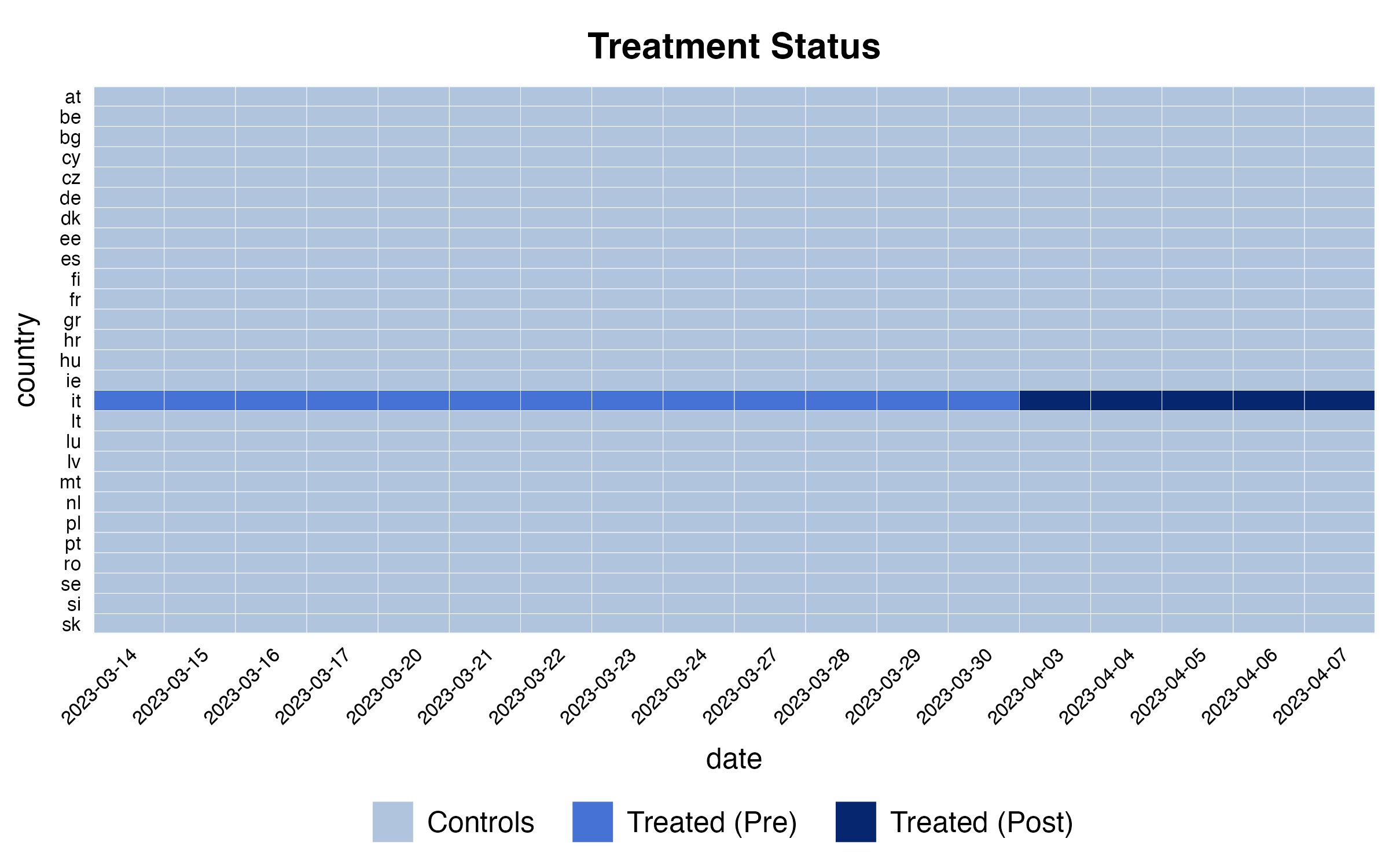}
\captionsetup{justification=justified, singlelinecheck=off} 
\caption*{\small \textit{Notes:} Panel structure of data sets used in analysis in Section \ref{sec:google-tor}. Note that the panel structure is equivalent for the \textit{(i) Google trends} and the \textit{(ii) Tor} (\textit{``standard''} and \textit{bridge relay}) users data set.}
\end{figure}

\end{document}